\documentstyle[12pt]{article}
\begin{document}

\newcommand{\dirac}[1]{/ \!\!\!#1}
\newcommand{\vgl}[1]{eq.(\ref{#1})}
\newcommand{\gv}{\gamma^5}
\newcommand{\gu}[1]{\gamma^{#1}}
\newcommand{\gd}[1]{\gamma_{#1}}
\newsavebox{\uuunit}
\sbox{\uuunit}
    {\setlength{\unitlength}{0.825em}
     \begin{picture}(0.6,0.7)
        \thinlines
        \put(0,0){\line(1,0){0.5}}
        \put(0.15,0){\line(0,1){0.7}}
        \put(0.35,0){\line(0,1){0.8}}
       \multiput(0.3,0.8)(-0.04,-0.02){12}{\rule{0.5pt}{0.5pt}}
     \end {picture}}
\newcommand {\unity}{\mathord{\!\usebox{\uuunit}}}
\newcommand{\half}{{\textstyle\frac{1}{2}}}
\newcommand{\dr}{\raise.3ex\hbox{$\stackrel{\leftarrow}{\partial }$}}
\newcommand{\dl}{\raise.3ex\hbox{$\stackrel{\rightarrow}{\partial}$}}

\begin{titlepage}
\begin{flushright} Preprint KUL-TF-92/41 \\
                   hepth@xxx/9301122\\
                   November 1992 \\
\end{flushright}

\vfill
\begin{center}
{\large\bf Hiding Anomalies }\\
\vskip 27.mm
{\bf F. De Jonghe$^{1,3}$, R. Siebelink$^{1,4}$, W. Troost$^{2,5}$ }\\
\vskip 1cm
Instituut voor Theoretische Fysica
        \\Katholieke Universiteit Leuven
        \\Celestijnenlaan 200D
        \\B-3001 Leuven, Belgium\\[0.3cm]
\end{center}
\vfill
\begin{center}
{\bf Abstract}
\end{center}
\begin{quote}
\small
Anomalies can be anticipated at the classical level without
changing the classical cohomology, by introducing extra degrees of freedom.
In the process, the anomaly does not quite disappear.
We show that, in fact, it is shifted to new symmetries
that come with the extra fields.
\vspace{2mm} \vfill \hrule width 3.cm
{\footnotesize
\noindent $^1$ Aspirant N.F.W.O., Belgium\\
\noindent $^2$ Bevoegdverklaard Navorser N.F.W.O., Belgium\\
\noindent $^3$ E-mail : Frank\%tf\%fys@cc3.kuleuven.ac.be \\
\noindent $^4$ E-mail : Ruud\%tf\%fys@cc3.kuleuven.ac.be \\
\noindent $^5$ E-mail : Walter\%tf\%fys@cc3.kuleuven.ac.be \\}
\normalsize
\end{quote}
\end{titlepage}

\parindent 5truemm
\parskip 0truemm

\section{Introduction}
The most interesting feature of the quantisation of anomalous gauge
theories lies in the fact that degrees of freedom that can be
gauged away classically become propagating at the quantum level. When
trying to quantise such theories there are at least two strategies
one can follow. The first one consists of first integrating out
the matter degrees of freedom. The result (see for examples
\cite{Polyakov,Sevrin}) is the induced action for the
gauge field. It is generically non-local, and is used thereafter as the
starting point for further quantisation, i.e.
integrated over to get the effective action.
The second possibility is to introduce some extra fields in the anomalous
theory. This line of developments started with
\label{???check WessZumino???}
Faddeev \cite{Faddeev}, who
showed that the presence of anomalies changes first-class constraints
(related to the presence of gauge-symmetries) into second-class
constraints. Later it was suggested in \cite{FadShat} that by introducing
extra degrees of freedom, one could keep the symmetries first-class, and
hence anomaly-free.
It was recognised that in a pathintegral approach the required new
variables arise very naturally when applying the Faddeev-Popov procedure.
In the case of a Yang-Mills model with chiral fermions for instance, it has
been shown \cite{Babelon} that the integral over the gauge group does not
factor out: the fermion measure has an (anomalous) dependence
on the gauge variables as well, effectively producing
the well-known Wess-Zumino action.

In a recent article \cite{Barcelona} these ideas have been implemented
in the
Batalin-Vilkovisky quantisationscheme. In this scheme anomalies arise as
the impossibility to find a local solution to the so called masterequation
$\half (W,W)=  \hbar \Delta W$ where W is the quantumaction i.e. the
classical action $S_{ext}$, extended with sources for the BRST
 transformation, some extra terms in case of open gauge algebra's,
 plus eventual (both renormalisation- and finite) counter\-terms $\hbar M_1 +
\hbar^2 M_2 + \cdots$. In a second step they add as many extra fields
to the
theory as there are anomalous symmetries.
By choosing their transformations
under the classical symmetries in a specific way,
 a local solution to the masterequation can actually
be found in this extended field space. This involves constructing a suitable
$M_1$-term, which in turn
makes these new fields dynamical on the quantum level.
In this approach, the anomaly apparently has disappeared, where the
price to pay is a (minor) change in the classical theory.

In this paper we
show how these extra fields can be introduced classically
via trivial systems, which
do not change the classical cohomology at all. This
approach makes clear that, at the same time as introducing new fields,
one also introduces extra symmetries. The extra
gaugesymmetries of the extra fields are properly taken into account. As
a result, we will show that the anomaly which seemed to have
disappeared, is in fact still present: it has merely been shifted,
or hidden, in these extra symmetries which are often ignored.

\section{Discussion}
In this section we will describe the method. We will formulate it
in the scheme of
Batalin and Vilkovisky (BV) for quantising gauge-theories \cite{BatVil}
\footnote{For a detailed account of BV, which includes the study of
anomalies in that scheme, we refer to \cite{boek,AnomBV}}, but will also
try to convey the central idea in less technical language.

In BV one starts from an extended action $S$, depending on the fields
$\phi^i$, on the ghost-fields $c^{a}$ for the gauge-symmetries of that
action, and on the anti-fields of all these fields, indicated with an
 asterisk,
which can be viewed as sources for the BRST transformations.
Assume that $S$ is a
solution of the classical master-equation $(S,S)=0$, i.e. is
BRST-invariant. A further requirement
is that it has
to satisfy the properness-condition, which means that ghosts have been
introduced for {\it all} the gauge-symmetries. In order to keep
 the symmetries
intact through the first quantum correction, one has to solve
the master-equation at order $\hbar$. This involves first regulating the
theory, and then calculating the operator $\Delta S$.
We assume that a Pauli-Villars (PV) regularisation is used to do this,
as described in \cite{AnomBV}. Anomalies are possible when the
 mass term chosen for the PV-fields does not maintain (all) the original
gauge symmetries. The result is always of
the form $$ (\Delta S)_{reg} = c^{a} {\cal A}_{a}. $$
If this quantity is the BRST-variation of something, than
the anomaly can be countered; if not, one speaks of a genuine
anomaly. We will always have in mind the latter case.

Now we propose to add trivial systems to $S$, one for every
anomalously broken gauge-symmetry:
$$ S \rightarrow S + \alpha^{*}_a d^a. $$
This may be viewed as follows. We introduced extra scalar degrees of
freedom $\alpha_a$. Since they do not occur in the original
action, they can always be shifted by an arbitrary amount
$\Lambda_a$. The $d^a$ are nothing but the ghosts for these extra
gauge-symmetries. Although these extra fields, and the entailing extra
symmetries, are completely trivial at this point (hence 'trivial systems'),
regularisation in the quantum theory will interfere with this.
The new degrees of freedom clearly do not change the classical theory, as
they are cohomologically trivial for the (new) cohomology-operator
 $(S,\cdot)$.

At this point, we are still free to specify how these newly introduced
fields should transform under the original symmetries related to the
$c-$ghosts. This choice can be encoded in the
extended action by doing a canonical transformation
\footnote{ Here we use 'canonical'
in the sense of anti-brackets. It is well-known that such a canonical
transformation does not change the classical cohomology.}.
Our approach then is
to choose this transformation such that a PV-mass term can be constructed
that {\it is} invariant under the $c$-symmetries.

Suppose one would like to have fields with transformations
$$ \delta_c \alpha^a = f^a(\phi,c,\alpha). $$ This can be achieved
simply by taking as generating fermion for the canonical transformation
\footnote{Remark that the field redefinitions of \cite{DamAlf} can be
implemented in BV in an analogous way.
If one wants to perform a field-redefinition given by
$ \phi_i = g_i(\alpha_j, \phi_i ')$, with the inverse relation
$ \phi_i ' = h_i(\alpha_j, \phi_i )$
then one should take as generating fermion
$ F = \phi'^{*}_i h^i + \alpha'^{*j} \alpha_j + d'^*_i d^i $}
$$ F = \unity - d_a^{' \ *} f^a(\phi,c,\alpha). $$
The fact that we used a canonical transformation guarantees that the
transformed extended action is still a solution of the classical
master-equation, i.e. the new action is still BRST-invariant with the
modified transformation rule.
 It is also important to remark that the extended action after the
transformation still contains terms with the $d$-ghosts. These are necessary
to ensure the properness of the action. It is clear that, if we use the
 extra $\alpha$-fields to construct a
PV mass-term that is invariant under the $c$-symmetries, it cannot also be
invariant under the $d$-symmetries, which shift $\alpha$. As a result
the anomalies will have been shifted to the new symmetries and
$$ (\Delta S)_{reg} = d^{\beta } {\cal B}_{\beta }. $$
It is only when one neglects the $d$-symmetries that one would conclude
 that there are no anomalies left.

To carry out the calculation one may still use the old
($c$ non-invariant) mass-term if one wishes to do so,
for technical or other reasons. The anomaly will still be left in the
$d$-symmetries, if a counterterm is added that matches
the interpolation between the two different regularisations (see
\cite{AnomBV}\cite{Wijzelf1}). This counterterm is completely fixed,
and involves an integration over a parameter that interpolates between
the two mass-terms.
In practise, it provides non-trivial dynamics for the variables
that were introduced as a trivial system.

\section{An example: W$_2$-gravity}

The action under consideration is
\begin{eqnarray}
S_{ext} = \int d^2x &&\!\!\!\! \frac{1}{2 \pi} ( \partial\varphi^i
\bar\partial\varphi^i
-h \  \partial\varphi^i \partial\varphi^i )\nonumber\\
&& +  \varphi_i^* \
\partial\varphi^i
c + h^* (\bar\partial c - h \partial c + c \partial h) + c^* (\partial c) c.
\label{extS}
\end{eqnarray}
We will only be concerned with the integral over the matterfields,
and define
\begin{eqnarray*}
\exp -\frac{1}{\hbar} \Gamma [h] &=&
\int {\cal D}\varphi \exp -\frac{1}{\hbar} S_{ext,\varphi^*=0}
\end{eqnarray*}
It is clear that we only have one-loop diagrams, so using PV-fields
the integral is completely regulated. We need to specify a massterm.
We choose the
obvious $S^{(0)}_{Mass} = - \frac{1}{2 \pi} M^2 (\varphi^i)^2$ : for the
covariant form of the action (\ref{extS}), it corresponds to keeping
the coordinate transformations anomaly-free. In that case we find the
well known Ward-identity
$$ \int d^2x  (\bar\partial c - h \partial c + c \partial h)
\frac {\partial \Gamma}{\partial h} = \frac{n \hbar}{12 \pi} \int d^2x \, c
\partial^3 h.  $$
Now we add the trivial system $\{\theta^*,d\}$.
We want to use it to shift the anomaly away from the $c$-symmetry
into the new $d$-symmetry, which amounts to constructing a $c$-invariant
massterm. Since $\delta_c S^{(0)}_{Mass} \sim c\partial (\varphi^i)^2$,
 we need a factor transforming as a density. Writing
$$ S^{(1)}_{Mass} = - \frac{1}{2 \pi} M^2 (\varphi^i)^2 e^\theta $$
we take
$$\delta_c \theta = \partial c + c \partial \theta .$$
The canonical transformation leading to this transformation
is generated by
$$ F = \unity - d'^* (\partial c + c \partial \theta). $$
The resulting extended action, in the new coordinates, is
\begin{eqnarray*}
S_{ext} = \int d^2x &&\!\! \frac{1}{2 \pi}(\partial\varphi^i
\bar\partial\varphi^i
-h \partial\varphi^i \partial\varphi^i) +  \varphi_i^*  \partial\varphi^i c
+ h^* (\bar\partial c - h \partial c + c \partial h) \\
      &+&\!\! c^* (\partial c) c
+ \theta^* (d + \partial c + c \partial \theta)
+ d^* (\partial d) c.
\end{eqnarray*}
This action is in fact the same (in simpler appearance)
as the action used in
\cite{Barcelona} except for the terms containing the $d$-ghost.
The origin of these terms was the extra trivial system,
and they are needed to assure that we have a proper extended action.

Next we calculate the anomaly due to the mattersector. As a
massterm we take
$S^{(\alpha)}_{Mass}=-\frac{1}{2 \pi} M^2 (\varphi^i)^2 e^{\theta \alpha}$
where $\alpha$ is a parameter interpolating between the conformal
($c$-) invariant
case ($\alpha = 1$), and the coordinate transformation invariant case
 ($\alpha = 0$).
We will not repeat here how to deduce the corresponding anomaly (follow
e.g. \cite{Hatsuda,Wijzelf1} for this application and \cite{Gilkey} for
the general method). The result for arbitrary $\alpha$ is
\begin{eqnarray*}
\Delta_{\varphi} S (\alpha) &=& n
\lim_{M\rightarrow \infty} \int d^2x \,\, \sqrt{g}
\frac{1}{4\pi} \left[ - \half \partial c + \half \alpha (
\partial c + d ) \right] \left[M^2 - \frac{1}{6} R\right]
\end{eqnarray*}
where $R$ is the Riemann scalar corresponding to the metric \\
 $g^{\mu\nu}=  e^{- \theta \alpha}    \left(\begin{array}{cc}
                                                    2h &  -1\\
                                                    -1 &   0
                                            \end{array} \right) $,
and $n$ is the number of scalars.\\
 The diverging term can be countered
by adding a local term to the action, since
$(S,\int d^2x\,e^{\alpha\theta})=\int d^2x\,\partial(e^{\alpha\theta} c)
  -\partial c +\alpha (d+\partial c)$, or alternatively by
considering several PV-fields with $\sum M^2=0$. The result is then
\begin{eqnarray*}
(\Delta_{\varphi} S)_{reg} (\alpha) &=& \frac{n}{12\pi} \int d^2x \,\,
\{\partial c - \alpha(\partial c + d)\}
 \{\partial^2h - \alpha (\partial \bar\partial \theta -
                                   \partial (h \partial \theta )) \}.
\end{eqnarray*}
Clearly, at $\alpha = 1$, the $c$-ghost disappears from this expression,
 and only the  $d$-ghost remains : this is one way to see that the
$c$-transformation anomaly is no longer present, but is replaced by a
$d$-transformation anomaly . Another way to see this is by looking at the
Ward identity
\begin{eqnarray*}
\int d^2x&&\!\!\!  (\bar\partial c - h \partial c + c \partial h)
\frac {\partial \Gamma}{\partial h} +  ( d + \partial c + c
\partial \theta ) \frac {\partial \Gamma}{\partial \theta }\\
&=& \frac{n \hbar}{12 \pi} \int d^2x \ \ d \{\partial^2h - \partial
\bar\partial
\theta + \partial (h \partial \theta ) \}
\end{eqnarray*}
where $\Gamma$ is the induced action computed with regulating mass term
 $S^{(\alpha)}_{Mass}$ at $\alpha=1$ :
\begin{eqnarray}
\exp -\frac{1}{\hbar} \Gamma [h,\theta] &=&
\int {\cal D}\varphi_{\{ \alpha=1 \}}
\exp -\frac{1}{\hbar} S_{ext,\varphi^*=0}.\label{gammaeen}
\end{eqnarray}
Actually, the conclusion above involves a choice: working with the
invariant $\theta$-dependent mass term, {\it and} not adding a finite
counterterm $M_1$.
If one prefers to work with the conventional massterm,
for example in Feynman diagrams, or because one wants
to use techniques with the usual OPE's, this can be done without changing
the resulting induced action if an
interpolating counterterm is added.
The expression for this counterterm can be obtained by considering that
$$\frac{\partial}{\partial \alpha} \log \det(-\Box +M^2 T(\alpha))=
   tr(\frac{M^2 \frac{\partial T(\alpha)}{\partial \alpha}}
                     {-\Box+M^2 T(\alpha)}) $$
with $\Box=\partial \bar\partial -\partial h \partial$, and applying
the technique of \cite{Diaz}.
This leads straightforwardly to the conjectured formula for the
counterterm $M_1$ given in \cite{AnomBV}
$$ M_1(\alpha) = - \frac{1}{2} \int^1_{\alpha}
             d\alpha \lim_{M^2\rightarrow \infty }
{\rm str}\ T^{-1}(\alpha) \frac{\partial T}{\partial \alpha}(\alpha)
e^{{\cal R}(\alpha)/M^2}, $$
where ${\cal R}$ is completely fixed.
Here, ${\cal R}(\alpha)=e^{\alpha \theta}\Box$
and in the present case the formula is exact since there is only one loop.
Therefore we also have, when using the mass term
$S^{(\alpha)}_{Mass}$, for any $\alpha$,
the following formula for the {\it same} induced action as in
eq.(\ref{gammaeen}):
\begin{eqnarray*}
\exp -\frac{1}{\hbar} \Gamma [h,\theta] &=&
\int {\cal D}\varphi_{\{ \alpha \}}
 \exp -\frac{1}{\hbar}( S_{ext,\varphi^*=0} + \hbar M_1(\alpha)), \\
M_1(\alpha)
    &=& \frac{n}{12\pi} \int d^2x \,\,
\half (1-\alpha^2)( \partial\theta \bar\partial\theta
     - h \, \partial\theta  \partial\theta)
 +(1-  \alpha) h \,\partial^2 \theta .
\end{eqnarray*}
The $\alpha$-independence can be checked for example by the fact that
$$ (\Delta S)_{reg}(\alpha)=(\Delta S)_{reg}(1)+(S,M_1(\alpha)). $$
In particular, for $\alpha=0$, we have the regularisation that does not
depend on the extra $\theta$-field. The matter field integral then
does not depend on $\theta$ either. In that case, the $M_1$
term completely exhibits the behaviour of the anomalous degree of freedom.
The variable $\theta$, originally introduced as part of a trivial
system, acquires dynamics by quantum corrections if we
require the c-symmetry -- including it's action on the extra field
$\theta$ -- to be anomaly-free. The $M_1$ term is actually the
Liouville action, which is not at all surprising if we realise that
$\delta_c \theta$ is precisely the residual reparametrisation
transformationlaw for the Liouville mode after a chiral gaugefixing.

Remark that in the quantumaction $W = S_{ext} + \hbar M_1(0)$ the
matterfields and the Liouville field $\theta$ can be treated on an
equal footing if we rescale
$ \theta \rightarrow \frac{1}{\sqrt \hbar} \theta ,
\theta^* \rightarrow \sqrt\hbar \theta^* $.
This redefinition clearly doesn't change the antibracketoperator and
we find (only retaining the matter- and Liouville transformations
under the c- symmetry) that
\begin{eqnarray*}
W = \int d^2x &&\!\!\!\! \frac{1}{2 \pi} ( \partial\varphi^i
\bar\partial\varphi^i -h \  \partial\varphi^i \partial\varphi^i )
+  \varphi_i^* \ \partial\varphi^i c \\
&&+ \frac{n}{24\pi} ( \partial\theta \bar\partial\theta
- h \, \{ \partial\theta  \partial\theta - 2 \sqrt\hbar \partial^2
\theta \} ) \\
&&+ \theta^* \{ \partial\theta c + \sqrt\hbar \partial c \}
\end{eqnarray*}
The backgroundcharge corrections to the
energy-momentum tensor, accompanied by quantumcorrected
transformationrules, appear naturally in this context. We also see that
their implementation in the BV scheme involves a slight subtelty: the
expansion in $\hbar$ contains actually half integral powers
$$ W = S_{ext} + \sqrt\hbar M_{\half} + \hbar M_1 + ... $$

We also wish to emphasize the close connection between regularisation
and the (finite) counterterm $M_1$. One can choose any regularisation
(any value of $\alpha$) , provided one adds the appropriate $M_1$.
Certainly, specifying only $M_1$ without fixing the regularisation is
insufficient : {\it any} of the above $M_1(\alpha)$ may be right.
Given a specific regularisation,
$M_1$ is determined by "external" requirements that one
imposes for the quantumtheory. In the above, this requirement was chosen
to be : absence of anomalies for the $c$-transformation.

The same ideas can easily be applied to other cases. Let us consider
chiral fermions in two dimensions, coupled to a gauge field.
One  starts from the action
$$ S[\psi,A] = \frac{1}{2\pi } \int d^2x \, \psi^t(\bar\partial-A)\psi, $$
and the classical symmetry is given by $ \delta \psi = c \psi $,
 and $ \delta A = \bar\partial c + [c,A] $.
This chiral gauge symmetry can be maintained at the quantumlevel if one
supplements the
theory with the group element $g$ as an extra field, not related
to $A$, with the
transformationrule $\delta g =(c +d)g $. Again, since this field is
not present in the action we started from, this entails extra
symmetries (with $d$-ghost)
that change $g$ arbitrarily. The fermions can only be regulated using
Pauli-Villars fields of both chiralities. One can again find
an invariant massterm
$$ S_{mass}^{(1)} [\psi,\xi] = - \frac{M}{4\pi} \int d^2x \, ( \psi^t \ g
\xi - \xi^t g^{-1} \psi). $$
Note that a similar construction was used also in  \cite{Slavnov}, in the
case of abelian chiral electrodynamics in four dimensions. In this paper
the importance of working with a regularised theory from the outset, was
also emphasized. \\
Technically, the calculation proceeds exactly as in \cite{Wijzelf1} (but
there the group element $g$ and the gauge field $A$ were related), so we
 do not repeat it here.
Interpolating to the chiral-non-invariant regularisation
characterised by the massterm
$$ S^{(0)}_{mass} [\psi,\xi] = - \frac{M}{4\pi}
\int d^2x\,(  \psi^t \xi - \xi^t \psi) $$
one picks up the counterterm
\begin{eqnarray*}
M_1 [g,A] &=& - S^{-}[g] - \frac{1}{2 \pi} \int d^2x \,\, tr \{ \partial g
\ g^{-1} A \} \\
S^{-}[g] &=&  \frac{1}{4 \pi } \int d^2x \,\, tr \{\partial g^{-1}
\bar\partial g\} - \frac{1}{12 \pi} \int d^3x \,\, \varepsilon^{\alpha
\beta \gamma } tr \{ g_{, \alpha} \ g^{-1}  g_{, \beta } \ g^{-1}
g_{, \gamma } \ g^{-1} \}
\end{eqnarray*}
This last term is of course the well-known Wess-Zumino-Witten action.
In the context of adding the (anomalous) quantum degrees of freedom,
 it has been proposed
before in \cite{Barcelona} or for the abelian case
in \cite{Braga}. As mentioned before, it is important to realise
that it's explicit expression is valid only when a specific
 regularisation is used. Our approach shows how
this counterterm originates from the fact that an
invariant regularisation exists.

\section{Summary}
Following the idea that the existence of an anomaly means
that extra degrees of freedom start propagating, we enlarge the set
of fields in the theory with an extra field for every anomalously
broken symmetry. Together with the extra fields, extra
gauge-symmetries are introduced in the theory, which act trivially.
We choose the transformation rules of the extra fields under the
original symmetries in such a way, that a regularisation exists that is
invariant under those symmetries. Adopting that regularisation,
those symmetries become anomaly-free. We find however that the anomalies
have not disappeared altogether : they have been shifted to the
extra symmetries, which at their introduction had a completely trivial
action.


\begin{thebibliography}{88}
\bibitem{Polyakov} A.M.~Polyakov, J.\ Mod.\ Phys.\ Lett.~A {\bf 2}
(1987)893.
\bibitem{Sevrin} K.~Schoutens, A.~Sevrin and P.~van Nieuwenhuizen,
in {\it Strings and Symmetries 1991}, World Scientific 1992, p558
\bibitem{Faddeev} L.D.~Faddeev, Phys. Lett. {\bf B145} (1984) p81.
\bibitem{FadShat} L.D.~Faddeev and S.L.~Shatashvili, Phys. Lett. {\bf B167}
(1986) p225.
\bibitem{Babelon} O.~Babelon,F.A.~Schaposnik and C.M.Viallet, Phys. Lett.
{\bf B177} (1986) p385.
\bibitem{Barcelona} J.~Gomis and J.~Paris, Universitat de Barcelona,
UB-ECM-PF 92/10.
\bibitem{BatVil}I.A. Batalin and G.A. Vilkovisky, Phys. Lett.
{\bf B102} (1981) 27; Phys. Rev. {\bf D28} (1983) 2567 [E: {\bf D30}
(1984) 508]; Nucl. Phys. {\bf B234} (1984) 106; J. Math. Phys.
{\bf 26} (1985) 172.
\bibitem{boek} W. Troost and A. Van Proeyen, {/it An introduction to
Batalin-Fradkin-Vilkovisky Lagrangian Quantisation},  Leuven
University Press, in preparation.
\bibitem{AnomBV} W.~Troost, P.~van~Nieuwenhuizen and A.~Van~Proeyen,
Nucl.\ Phys.~{\bf B333}(1990)727.
\bibitem{DamAlf} Alfaro and Damgaard, Ann. Phys. {\bf 202}1990, p398
and also CERN-TH-6455/92.
\bibitem{Wijzelf1} F. De Jonghe, R. Siebelink, W. Troost, Phys. Lett. {\bf
B288} (1992) p47.
\bibitem{Hatsuda}
M. Hatsuda, W. Troost, P. van Nieuwenhuizen
 and A. Van Proeyen, Nucl. Phys. {\bf B335} (1990) 166.
\bibitem{Gilkey} P.B.Gilkey, Proc. Sympos. Pure Math {\bf 27} Amer.
Math. Soc.(1975)265;\\ J.\ Diff.\ Geom.~{\bf 10}, 601 (1975).
\bibitem{Diaz} A. Diaz, W. Troost, P. van Nieuwenhuizen
 and A. Van Proeyen, Int. J. Mod. Phys. {\bf A4}  (1989) 3959.
\bibitem{Slavnov} S.~A.~Frolov and A.~A.~Slavnov, Phys. Lett. {\bf
B269} (1991) p377
\bibitem{Braga} N.~R.~F.~Braga and H.~Montani, Phys. Lett. {\bf
B264} (1991) p125
\end{thebibliography}
\end{document}